\documentstyle[12pt]{article}          
\newcommand{\be}{\begin{equation}}          
\newcommand{\en}{\end{equation}}          
          
\newcommand{\Sc}{Schr\"odinger}          
\newcommand{\E}{equation}          
 \newcommand{\al}{\alpha }          
 \newcommand{\vfi}{\varphi }          
        
\input{amssym.def}

\begin{document}          

\hfill quant-ph/9904009

\begin{center}          
\noindent          
{\Large \bf New possibilities for supersymmetry breakdown          
in quantum mechanics and second order irreducible Darboux         
transformations          
}\\[5mm]          
          
\noindent          
{\large Boris F. Samsonov} \\          
{\small \it Department of Quantum Field Theory, Tomsk State          
University, 36 Lenin Ave., \\634050 Tomsk, Russia,          
email: samsonov@phys.tsu.ru}          
\end{center}          
          
\vspace{3mm}          
\small          
\noindent          
{\bf Abstract.}          
         
New types of irreducible second order Darboux       
transformations          
for the one dimensional Schr\"o\-di\-nger equation  are          
described. 
The main feature of such transformations is that the          
transformation functions have the eigenvalues grater then the          
ground state  energy of the initial (or reference) Hamiltonian.          
When such a transformation is presented as a chain of two first          
order transformations, an intermediate potential is singular and          
therefore intermediate Hamiltonian can not be Hermitian while          
the final potential is regular and the final Hamiltonian is          
Hermitian. Second derivative supersymmetric quantum mechanical          
model based on a transformation of this kind exhibits properties inherent       
to models with exact          
and broken supersymmetry at once.         
         
         

\vspace{3mm}

\normalsize          
         
{\bf 1.} Supersymmetric quantum mechanics (SUSY QM) that has been          
introduced to illustrate problems of supersymmetry breakdown in          
quantum field theories finds now numerous applications in          
different fields of theoretical and mathematical physics (for a          
review see \cite{1}). It is well known that the supersymmetry          
may be either exact or broken. In the case of the broken          
supersymmetry the entire spectrum of the supehamiltonian is          
degenerate while in the case of the exact one its vacuum state          
is nondegenerate. We would like to stress that other          
possibilities exist as well \cite{2,3} that may have          
applications in quantum field theories \cite{2}. 
         
In the conventional SUSY QM \cite{1} supercharges are built of          
first order Darboux transformation operators \cite{An1}.          
Higher order Darboux transformation operators are involved          
in a higher derivative supersymmetry \cite{An2,tmf}. In the          
simplest case we have a second order derivative supersymmetry          
with the supercharges built of the second order Darboux          
transformation operators. It is known that such a supersymmetry          
may be either reducible or not \cite{An3}. The concept of          
complete reducibility is introduced as well \cite{3,rev}. This          
concept is based on a theorem that establishes the equivalence          
between an $N$-order Darboux transformation and a chain of $N$          
first order Darboux transformations \cite{rev}.          
          
Every chain of $N$ first order Darboux transformations          
creates a          
chain of exactly soluble Hamiltonians $h_0\to h_1\to \ldots \to          
h_N$. We suppose that $h_0$ and $h_N$ are Hermitian in a          
Hilbert space and admit unique self-adjoint extensions (i.e.          
they are essentially self-adjoint). To satisfy this condition          
the potentials $V_i(x)$, $h_i=-\partial _x^2+V_i(x)$,
$\partial _x^2=\partial _x\partial _x$, $\partial _x=d/dx$,
$i=0,N$ should be realvalued and free of singularities in their common          
domain of definition $(a,b)$ where $a$  or $b$ or both may be          
equal to $+ \infty $ or $- \infty $ . Any chain is called {\it       
reducible} \cite{An3}         
if all intermediate potentials $V_1(x)$, $\ldots $, $V_{N-1}(x)$         
are realvalued functions defined in $(a,b)$.      
The $N$th order Darboux transformation which is equivalent to       
the whole chain is called reducible as well. When at least one of       
the intermediate potentials is a complexvalued function the       
chain and the corresponding $N$th order transformation are       
called {\it irreducible}.      
Some reducible         
chain is called {\it completely reducible} \cite{2} if all these         
potentials are free of singularities in $(a,b)$. The       
corresponding $N$th order transformation is called completely       
reducible as well.         
         
{\bf 2.} It can be shown that every $N$th order Darboux         
transformation is equivalent to the resulting action of a chain         
of well-defined transformations of order less or equal two.         
There is a marvellous vast literature devoted to the analysis of 
first order transformations (see e.g. \cite{1}). 
Second order transformations are not explored in the same 
details. The main purpose of this letter is to fulfil this gap
and give an analysis of second order  transformations.
         
Let $L$ be a second order Darboux transformation operator         
defined as an operator that intertwines two well-defined         
Hamiltonians $h_0$ and $h_2$, $Lh_0=h_2L$. Every operator of         
this type may be presented in a compact form as follows         
\cite{rev}: $L\psi =W^{-1}(u_1,u_2)W(u_1,u_2,\psi )$ where          
$W$ stands for the usual symbol of a Wronskian and the         
functions $u_1(x)$ and $u_2(x)$ called          
{\it transformation functions} are eigenfunctions of $h_0$,          
$h_0u_{1,2}=\al _{1,2}u_{1,2}$ which are not supposed to         
satisfy any boundary conditions. When supercharge operators $Q$         
and $Q^+$ are built in terms of $L$ and $L^+$ where $L^+$ is         
Laplace adjoint to $L$         
$$Q=      
\left( \begin{array}{cc} 0 & L^+ \\ 0 & 0\end{array} \right),      
\quad       
Q^{+}=\left(\begin{array}{cc} 0 & 0 \\ L & 0\end{array}         
\right)  $$         
then these operators together with the superhamiltonian          
${\cal H}={\rm diag}(h_0,h_2)$ close a second order         
superalgebra:         
$Q^2=(Q^+)^2=0$,          
$QQ^++Q^+Q=({\cal H}-\al _1{\cal I})( {\cal H}-\al _2{\cal I})$         
where ${\cal I}$ is the unit $2\times 2$ matrix. The new        
potential $V_2(x)$ is defined by the initial potential $V_0(x)$        
and the transformation functions        
$V_2(x)=V_0(x)-2[\log W(u_1,u_2)]^{\prime \prime }$ where the prime stands for the derivative with respect to $x$.
      
The operator $L$ can always be presented as a superposition       
of first order operators      
$L=L^{(1)}L^{(2)}$ where       
$L^{(1)}=-\partial _x+(\log u_1)^\prime $,        
  $L^{(2)}=-\partial _x+(\log v)^\prime $, $v= L^{(1)}u_2$. The       
intermediate potential of such a chain of transformations is       
defined by the function $u_1$:       
$V_1(x)=V_0(x)-2(\log u_1)^{\prime \prime }$.       
       
In the conventional SUSY QM \cite{1} transformation functions         
are supposed to be such that their eigenvalues are subject to         
the condition $\al _{1,2}\le E_0$ where $E_0$ is the ground         
state energy when $h_0$ has a discrete spectrum and it is the         
low bound of the continuum spectrum when the discrete spectrum         
is absent. They also may be chosen such that $\al _1=E_0$, $\al         
_2=E_1$ where $E_1$ is the energy of the first excited state.         
These choices correspond to the usual conception of the         
supersymmetry breakdown in quantum mechanics when the vacuum         
state of ${\cal H}$ is nondegenerate for the exact         
supersymmetry and it is two fold degenerate for the broken         
supersymmetry. In the latter case the whole spectrum of ${\cal         
H}$ is two fold degenerate.         
        
{\bf 3.} In what follows we shall denote by $E_k$ and $\psi _k$
$k=0,1,\ldots $         
the discrete spectrum eigenvalues and eigenfunctions of the         
Hamiltonian $h_0$ respectively. We will suppose for simplicity that         
 the whole spectrum of $h_0$ is discrete.       
        
We shall prove below that $\al _{1,2}$ may be chosen such that         
$E_{k+1}\ge \al _2>\al _1\ge E_k$. In this case the         
transformation functions $u_{1,2}$ has nodes in $(a,b)$ and         
consequently intermediate potential $V_1(x)$ has singularities         
in $(a,b)$. Nevertheless there exists such a choise of the         
transformation functions that the final potential $V_2(x)$ is         
free of singularities. We obtain thus a simplest 
irreducible chain of Darboux transformations.       
       
Such a transformation either deletes one or two energy levels        
($E_k$ or $E_{k+1}$ or both) or creates one or two new energy        
levels disposed between $E_k$ and $E_{k+1}$ ($\al _1$ or $\al        
_2$ or both). The SUSY QM based on this transformation has the        
properties of the theories with the exact and broken        
supersymmetry at once. The ground state of ${\cal H}$ is two        
fold degenerate and in the middle of the spectrum of ${\cal H}$        
there exist one or two nondegenerate energy levels.

{\bf 4.} Let us establish now conditions for the potential        
$V_2(x)$ to be free of singularities in the interval $(a,b)$.     
For        
this purpose it is sufficiently to analyse the second order        
Wronskian        
$W(u_1,u_2)$ as a function of $x$ and find the conditions when        
it is free of zeros in $(a,b)$.        
       
For the sake of definiteness we will consider the full real axis        
$\Bbb R =(-\infty , \infty )$ as the interval $(a,b)$ and        
suppose the potential $V_0(x)$ to be confining,
i.e. $|V_0(x)|\to        
\infty $ as $|x|\to \infty $. We also assume the potential       
$V_0(x)$        
to be sufficiently smooth function in $\Bbb R$ (e.g. $V_0(x)\in        
C^\infty _{\Bbb R} $) and bounded from below. In this case the        
operator $h_0=-\partial _x^2+V_0(x)$ initially defined on a       
set        
of infinitely differentiable functions with a compact support        
which is dense in the Hilbert space of functions defined over        
$\Bbb R$ and square integrable with respect to the Lebesgue measure has a        
self adjoint closure that we shall denote $h_0$ as well.        
Moreover, $h_0$ has only a discrete spectrum $E=E_k$ with the        
eigenfunctions $\psi =\psi _k$, $k=0,1,\ldots $.       
       
It is not difficult to see that every eigenfunction of $h_0$,       
$\psi _E$, with $E$ such that $E_{k+1}>E>E_k$,       
$k=0,1, \dots $ may have on $\Bbb R$ either $k+1$ nodes or $k+2$ ones. Moreover, If $\psi _E $ has $k+2$ nodes then       
$|\psi _E(x)|\to \infty $ as $|x|\to \infty $. If $\psi _E$ has       
$k+1$ nodes then we have two possibilities: a) $\psi _E \to 0$       
as $x\to \infty $ (or equivalently as $x\to -\infty $) and b)       
$|\psi _E|\to \infty $ as $|x|\to \infty $. 
In the first case       
we refer $\psi _E$ as the function with a zero asymptotic at the right infinity (or equivalently at the left infinity) and       
in the second case as the one with a growing asymptotic at both infinities. If $\psi _E$ is a zero asymptotic function then       
it has just $k+1$ nodes in $\Bbb R$. These assertions are       
direct implications of the well-known Sturm oscillator theorem     
(see e. g.       
\cite{Ber}).      
      
Our analysis shows that if the transformation functions $u_1$       
and $u_2$, $h_0u_{1,2}=\al _{1,2}u_{1,2}$ are chosen such that       
$E_{k+1}\ge \al _2>\al _1>E_k$, $k=0,1,\ldots $ and $u_1(x)$       
has $k+2$ nodes, $u_2(x)$ has $k+1$ nodes on $\Bbb R$ then       
$W(u_1,u_2)=W(x)$ is free of zeros on $\Bbb R$. Really. We note      
first of all that because of the conditions imposed on $u_1$      
and $u_2$ they have simple and alternating zeros i.e. between      
any two consecutive zeros of one of them there is exactly one      
zero of the other. The total number of nodes of the functions      
$u_1$ and $u_2$ is odd and equal 2k+3. Let $x_0$, $\ldots $,      
$x_{2k+2}$ be the zeros of the functions $u_1(x)$ and $u_2(x)$.     
Since $W^{\prime }(x)=(\al _1-\al _2)u_1(x)u_2(x)$ where the      
use of the Schr\"odinger equation is made the points $x_0$,      
$\ldots $, $x_{2k+2}$ are the points of local minima and      
maxima of $W(x)$. It then follows that this function is      
monotone in every interval $[x_j,x_{j+1}]$, $j=0,1,\ldots $ and      
the points $x_0$ and $x_{2k+2}$ are both either maxima or      
minima. It is not difficult to see that the sign of the      
function     
$W(x)=u_1(x)u^\prime _2(x)- u_2(x)u^\prime _1(x)$ is the same     
for all $x=x_0,\ldots x_{2k+2}$ and $W(x)\ne 0$      
$\forall x\in [x_0,x_{2k+2}]$.     
It remains now to analyse the behaviour of the function $W(x)$      
for $x<x_0$ and $x>x_{2k+2}$.     
    
The sign of the functions      
$u_1(x)$ and $u_2(x)$ is without importance. Therefor without     
loss of generality we can always choose $u_1(x)$ and $u_2(x)$     
such that $u_{1,2}(x)\ge 0$ for $x>x_{2k+2}$. In this case     
$u_1(x_{2k+2})=0$, $u_2(x_{2k+2})>0$,    
 $u_1^{\prime }(x_{2k+2})>0$. We conclude that     
$W(x_{2k+2})<0$ and $W^{\prime \prime }(x_{2k+2})<0$. This     
means that $x_0$ and $x_{2k+2}$ are the points of local maxima     
of $W(x)$. Taking into account the fact that $W(x)$ is monotone     
for $x> x_{2k+2}$ and $x<x_0$     
(since $W^{\prime }(x)\ne 0$ for all these $x$) and negative we     
find that $W(x)\ne 0$ for all $x\in \Bbb R$.    
    
 {\bf 5.} It is clear from the above considerations that if    
$u_1(x)$ has $k+1$ nodes and $u_2(x)$ has $k+2$ nodes then    
under the same assumptions that above the function $W(x)$ is    
positive and riches the local maxima at $x=x_0$ and    
$x=x_{2k+2}$. Further, the function $u_1(x)$ is not square    
integrable on $\Bbb R$ because of the condition    
$E_{k+1}\ge \al _2>\al _1>E_k$. The function $u_2(x)$ is not    
square integrable for all $\al _2\in [E_{k+1},E_k)$ since it is    
assumed to have $k+2$ nodes. It follows then that    
$|u_2(x)|\to \infty $ as $|x|\to \infty $ and $|u_1(x)|$ tends    
to infinity in one of the infinities at least. This means that    
$|W(x)|$ tends to infinity in one of the infinities at least    
and $|W(x)|$ has one node at least.   
   
We could avoid such a behaviour of $W(x)$ if $|W(x)|$ would    
decrease as $|x|\to \infty $ instead of increase. For such a    
behaviour of $|W(x)|$ the function $u_1(x)$ should be square    
integrable over $\Bbb R$. (Note that the case when $u_2(x)$ is    
square integrable was considered above.) I could not prove that    
for an arbitrary potential $V_0(x)$ the Wronskian $W(x)$ tends in    
this case to zero as $|x|\to \infty $. Nevertheless, I can    
indicate a wide class of potentials for which the condition    
$W(x)\to 0$ as $|x|\to \infty $ holds. In particularly, all potentials satisfying the condition   
$$\int _{-\infty }^{+\infty }|xV_0(x)|dx<\infty $$   
(scattering potentials)
are of this type. As to confining potentials it can be proven    
that when the potential $V_0(x)$ is such that    
$$   
\int _{-\infty }^\infty      
\left| \frac{V_0^\prime }{V_0^{5/4}} \right|^2 dx<\infty ,\quad      
\int _{-\infty }^\infty      
\frac{|V_0^{\prime \prime }|}{|V_0|^{3/2}}  dx<\infty .      
$$   
then $W(x)\to 0$ as $|x|\to \infty $. The proof of this    
assertion is based on the asymptotic behaviour of the solutions    
of the Schr\"odinger \E \ for such potentials which is known (see    
e.g. \cite{Ber}) and it is omitted here. This imply 
another possibilities for the choice of the transformation   
functions. 
If they are chosen such that   
$E_{k+1}\ge \al _2>\al _1=E_k$, $u_1(x)=\psi _k(x)$, and   
$u_2(x)$ has $k+1$ nodes then $W(u_1,u_2)\ne 0$   
$\forall x\in {\Bbb R}$. The proof can easily   
been obtained with the help of the asymptotic behaviour of   
$u_1(x)$ and $u_2(x)$.   
   
{\bf 6.} We have formulated conditions that are   
sufficient to impose on the transformation functions to obtain    
a regular potential of the transformed \Sc \ \E \ with the potential $V_2(x)$. It can be   
shown with the aid of the known asymptotic of the solutions of   
the initial \Sc \ \E \ that $V_2(x)\to V_0(x)$ as   
$|x|\to \infty $. Moreover, the knowledge of this asymptotic  
allows us to find all solutions of the transformed \Sc \  equation that belong  
to the Hilbert space $L_2({\Bbb R})$. This analysis is based on  
the following affirmations. 
 
The spectrum of the Hamiltonian $h_N$ related with $h_0$ by the  
$N$th order Darboux transformation coincides with the spectrum  
of $h_0$ with the possible exception of a finite number of  
discrete levels defined by the choice of the transformation  
functions. The level $E=\al $ is absent in the spectrum of  
$h_N$ if and only if $u_\al \in {\rm Ker}L$,  
$h_0u_\al =\al u_\al $ and $u_\al \in L_2({\Bbb R})$. The level  
$E=\al $ is created in the spectrum of $h_N$ if and only if  
$v_\al \in {\rm Ker}L^+$,  
$h_Nv_\al =\al v_\al $ and $v_\al \in L_2({\Bbb R})$. The space  
${\rm Ker}L^+$ has the basis \cite{rev} $v_1(x),\ldots ,v_N(x)$, 
$v_j=W^{(j)}(u_1,\ldots u_N)/W(u_1,\ldots u_N)$,
$h_Nv_j=\al _jv_j$ 
where $W^{(j)}(u_1,\ldots u_N)$ is the Wronskian of order 
$N-1$ built of the functions $u_1,\ldots ,u_N$ except for the  
function $u_j$. 
 
The direct implication of this assertion is that the  
functions $\vfi_k=L\psi _k$ are square integrable over $\Bbb R$ 
for all $\psi _k\ne u_j$. To find all square integrable  
solutions of the transformed \Sc \ \E \ it remains now to  
analyse the functions 
$v_{1,2}=u_{2,1}/W(u_1,u_2)\in {\rm Ker}L^+$. This  
analysis is possible because of the known asymptotic of the  
solutions of the initial \Sc \ \E . We have obtained the  
following results. 
 
If the transformation functions $u_{1,2}$ are such that  
$E_{k+1}>\al _2>\al _1>E_k$ and have growing asymptotic on both  
infinities then $v_{1,2}\in L_2({\Bbb R})$ and the set 
$\left\{v_1,v_2,\vfi _n=L\psi _n, n=0,1,\ldots \right\}$ is  
complete in  
$ L_2({\Bbb R})$. (Hamiltonian $h_2$ has two additional energy  
levels $E=\al_1, \al _2$ with respect to $h_0$.) 
If the  
function $u_2$ has zero asymptotic then  
$v_2\notin L_2({\Bbb R})$, $h_2$ has a single additional  
energy level $E=\al _1$, and the set  
$\left\{v_1,\vfi _n=L\psi _n, n=0,1,\ldots \right\}$ forms a  
basis in $ L_2({\Bbb R})$. 
 
If $\al_2 =E_{k+1}$, $u_2=\psi _{k+1}$, and $u_1$ has a  
growing asymptotic at both infinities then the level  
$E=E_{k+1}$ is absent in the spectrum of $h_2$ and the level 
$E=\al _1$ is created. The set  
$\left\{v_1,\vfi _n=L\psi _n, n=0,1,\ldots ; n\ne k+1\right\}$
is complete in $ L_2({\Bbb R})$.
 
If $\al _1=E_k$, $u_1=\psi _k$, and $u_2$ has a growing  
asymptotic at both infinities then the level $E=E_k$ is absent  
in the spectrum of $h_2$ and the level $E=\al _2$ is created. 
The basis in $ L_2({\Bbb R})$ is formed by the set 
$\left\{v_2,\vfi _n=L\psi _n, n=0,1,\ldots ; n\ne k\right\}$. 
When $u_2$ has a zero asymptotic then we have only a deletion of  
the level $E=E_k$ and the basis is formed by the set 
$\left\{ \vfi _n=L\psi _n, n=0,1,\ldots ; n\ne k\right\}$. 
Finally, if $\al _2=E_{k+1}$, $\al_1=E_k$,
and $u_{1}=\psi _k $, $u_2=\psi _{k+1}$
 then both levels  
$E=E_{k}$ and $E=E_{k+1}$ are absent in the spectrum of $h_2$ and  
the position on the other levels is unchanged. 
The set  
$\left\{ \vfi _n=L\psi _n, n=0,1,\ldots ; n\ne k, k+1\right\}$  
is complete in $ L_2({\Bbb R})$. This possibility has been earlier indicated by Krein \cite{Kr}.

{\bf 7.} As a final remark we note that the possibility to use the transformation functions with the eigenvalues higher then the ground state energy of $h_0$ has recently been noted in \cite{FHM} without any analysis.

The financial support from the RFBR and the ministry of education of Russia is gratefully acknowledged.

\end{document}